\documentclass[prb,twocolumn,showpacs,preprintnumbers,amsmath,amssymb]{revtex4}

\usepackage{amsmath,amssymb,amsfonts}
\usepackage[dvips]{graphicx}
\usepackage[latin1]{inputenc}
\usepackage{subfigure}
\begin{document}

\title{Vibrationally Mediated Control of Single Electron Transmission in Weakly Coupled Molecule-Metal Junctions}

\author{Thomas Olsen}
\email{tolsen@fysik.dtu.dk}
\author{Jakob Schi{\o}tz}

\affiliation{Danish National Research Foundation's Center for Individual
	Nanoparticle Functionality (CINF),
	Department of Physics,
	Technical University of Denmark,
	DK--2800 Kongens Lyngby,
	Denmark}

\date{\today}

\begin{abstract}
We propose a mechanism which allows one to control the transmission of single electrons through a molecular junction. The principle utilizes the emergence of transmission sidebands when molecular vibrational modes are coupled to the electronic state mediating the transmission. We will show that if a molecule-metal junction is biased just below a molecular resonance one may induce the transmission of a single electron by externally exciting a vibrational mode of the molecule. The analysis is quite general but requires that the molecular orbital does not hybridize strongly with the metallic states. As an example we perform a density functional theory (DFT) analysis of a benzene molecule between two Au(111) contacts and show that exciting a particular vibrational mode can give rise to transmission of a single electron. 
\end{abstract}

\maketitle
Several experiments have established that vibrational excitations can have a significant effect on the \textit{I}-\textit{V} characteristics of single-molecule junctions \cite{park, yu, pasupathy, sapmaz, leturcq}. In particular, the emergence of peaks in the differential conductance corresponding to vibrational frequencies, shows that tunneling electrons interact with certain vibrational states of the molecule, possibly providing a means for controlling the transmission of electrons. A considerable amount of theoretical work have been dedicated to elucidating the effect of phonons on electronic transport in mesoscopic systems. Analysis of model Hamiltonians \cite{wingreen, mii, braig, mitra2} have given qualitative insight into the interaction of tunneling electrons with molecular vibrations, while DFT based studies in conjunction with a non-equilibrium Green function approach show quantitative agreement with experiments \cite{galperin, frederiksen, Kristensen}. Most efforts so far have been directed towards the influence of vibrations on transmission functions and \textit{I}-\textit{V} characteristics. In the present paper we will take a slightly different point of view and partition the electronic transmission function into pieces that involve different vibrational excitations. We then show that the molecular junction may be put in a configuration where the vibrationally excited molecule allows the transmission of a single electron, while transmission is forbidden in the vibrational ground state. Controlling the vibrational state of the molecule, e.g. by means of a laser, then implies control of single electron transmission.

\begin{figure}[tb]
	\includegraphics[width=6.0 cm]{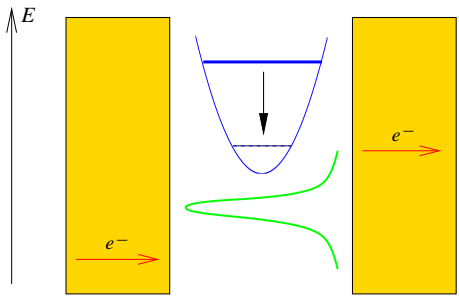}	
\caption{A molecule between two metal contacts is represented by a resonant state (for example the lowest unoccupied molecular orbital) and a vibrational potential. If the molecule is initially vibrationally excited, an electron below the resonance may tunnel through the molecule by absorbing a quantum of vibration. When the molecule is initially in its vibrational ground state, transmission is not possible.}
\label{fig:res}
\end{figure}

The system under consideration is a molecule sandwiched between two metallic leads. We assume that there is a single unoccupied molecular state which obtains a finite lifetime due to hybridization with metallic states. This state will be referred to as the resonance and its position may be tuned by applying a gate voltage. If a bias voltage is applied to the contacts and the resonance is positioned in the bias window, electrons may tunnel through the resonance into the downstream contact and one will observe a current. If the molecule is weakly interacting with the metal such that the resonance is well localized in energy, one can apply a gate voltage which situates the resonance above the upstream chemical potential and no current will be observed. However, if the resonance couples to molecular vibrations, an off-resonant enhancement of transmission known as transmission sidebands may be observed \cite{wingreen}. In particular, if the molecule is initially vibrationally excited, off-resonant electrons below the resonance may tunnel though the molecule by absorbing a vibrational quantum of energy. This is illustrated in Fig. \ref{fig:res}. Since transmission is only allowed combined with a downward vibrational transition, only one or a few electrons may tunnel through the contact and the transmission channel will be closed once the vibrational mode reaches the ground state.

Inspired by these considerations, we perform a quantitative analysis based on the model Hamiltonian \cite{wingreen, galperin_review}
\begin{align}\label{H}
H&=\varepsilon_0c_a^{\dag}c_a + \sum_i\hbar\omega_ib_i^{\dag}b_i+\sum_i\lambda_ic_a^{\dag}c_a(b_i^{\dag}+b_i)\\
&+\sum_k\epsilon_{Lk}c_{Lk}^{\dag}c_{Lk}+\sum_k\Big(V_{Lk}c_a^{\dag}c_{Lk}+V_{Lk}^*c_{Lk}^{\dag}c_a\Big)\notag\\
&+\sum_k\epsilon_{Rk}c_{Rk}^{\dag}c_{Rk}+\sum_k\Big(V_{Rk}c_a^{\dag}c_{Rk}+V_{Rk}^*c_{Rk}^{\dag}c_a\Big),\notag
\end{align}
where $c_a^\dag$ is the creation operator for the lowest unoccupied molecular orbital (LUMO), $c_{Lk}^\dag$ and $c_{Rk}^\dag$ are creation operators for metallic states in the left and right lead respectively and $b_i^\dag$ are creation operators for the vibrational normal modes of the molecule with frequencies $\omega_i$. Thus, the electronic states of the left and right contacts are coupled through the molecular resonance which is coupled to molecular vibrations with coupling strengths $\lambda_i$. We impose the wide band limit in which the contact density of states is constant in the region of the resonance and the resonance hybridization with metallic states is determined by the parameters 
\begin{align}\label{gamma}
\Gamma_{L}=2\pi\sum_k|V_{Lk}|^2\delta(\varepsilon_0-\epsilon_{Lk}),
\end{align}
and a similar expression for the coupling to the right lead $\Gamma_R$. Without the vibrational coupling, the resonance spectral function would then be a Lorentzian with full width at half maximum given by $\Gamma=\Gamma_L+\Gamma_R$. We will be interested in the regime $\Gamma,\;k_BT\ll\hbar\omega_i$, but we will not restrict ourselves to the classical limit $\Gamma\ll k_BT$ where the contact current can be expressed in terms of rate equations \cite{braig,mitra1,mitra2}. Instead, we consider the transmission matrix $T_n(\varepsilon_i,\varepsilon_f)$ for an electron with initial state energy $\varepsilon_i$ and final state energy $\varepsilon_f$.
Within scattering theory, the transmission matrix can be expressed in terms of a two-particle Green function \cite{wingreen} which can be evaluated exactly in the wide band limit \cite{olsen2}. 
The subscript $n$ refers to the initial state of the oscillator and integrating out the final state energy, one obtains the transmission probability $T_{nm}(\varepsilon_i)$ that an electron with initial state energy $\varepsilon_i$ is transmitted while the molecule makes the vibrational transition $n\rightarrow m$. 

\begin{figure}[tb]
	\includegraphics[width=7.0 cm]{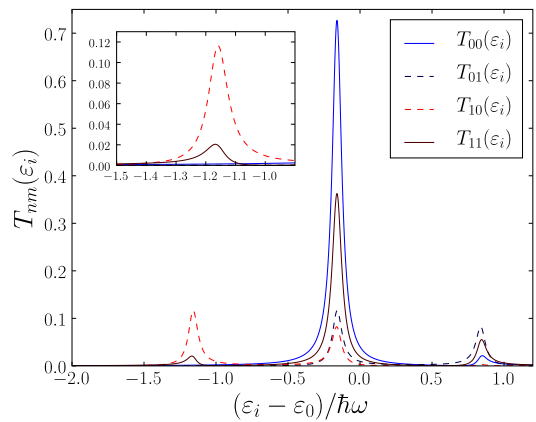} 
\caption{Transmission probabilities calculated from \eqref{H} as a function of incoming electron energy with $\Gamma_L=\Gamma_R=0.04\hbar\omega$ and $\lambda=0.4\hbar\omega$. Below the resonance energy $\varepsilon_0$ the ground state transmission functions $T_{00}(\varepsilon_i)$ and $T_{01}(\varepsilon_i)$ essentially vanish. The insert shows the lower sideband, where stimulated emission of a vibrational quantum $T_{10}(\varepsilon_i)$ is the dominating transmission channel.}
\label{fig:transmission}
\end{figure}
In the appendix we have calculated the transmission matrix in the ground and first excited state and in Fig. \ref{fig:transmission} we show the transmission probability corresponding to four different vibrational transitions of the molecule. It shows that incoming electrons with energies below $\varepsilon_0-\hbar\omega$ have a vanishing probability of transmission ($\sum_nT_{0n}$) when the molecule is in its vibrational ground state. This means that if a bias voltage is applied such that the Fermi level of the upstream contact is at $\varepsilon_F=\varepsilon_0-\hbar\omega$ no current will be observed when the molecule is in its vibrational ground state. If the molecule is vibrationally excited, e.g. by means of a IR laser, transmission becomes possible through the low lying vibrational sideband ($T_{11}$ and $T_{10}$). However, the first electron which is transmitted through the $T_{10}$ channel induces a transition to the vibrational ground state and thus closes the transmission channel completely. Hence, the net effect is that applying a short laser pulse to the molecular contact can induce the transmission of a few electrons.

\begin{figure}[b]
	\includegraphics[width=5.0 cm]{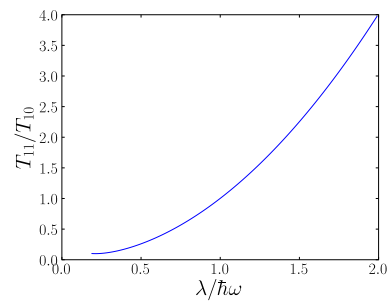}
\caption{Ratio of elastic and inelastic transmission probabilities in the first vibrationally excited state. At weak coupling the elastic transmission $T_{11}$ goes to zero indicating that a vibrational excitation results in only a single electron being transmitted.}
\label{fig:ratio}
\end{figure}

The distribution of the number of electrons being transmitted as a result of a vibrational excitation depends on the ratio $T_{11}/T_{10}$ shown in Fig. \ref{fig:ratio} as a function of the vibrational coupling. For small coupling parameters the ratio approaches zero and the first electron to be transmitted is therefore highly likely to induce a vibrational decay and close the channel.
For $\lambda\ll\hbar\omega$ we thus have $T_{11}\ll T_{10}$ which is needed for single electron transmission. One might worry that the inelastic transmission amplitude may become too small for anything to happen in this case. However, the absolute amplitude of sideband transmission can easily be small if the vibrational lifetime of the molecule is long. For physisorbed molecules the lifetime of a vibrational state is typically on the order of nanoseconds, whereas for example with Au(111), we can use the density of states to estimate that $\sim40$ electron hit each surface atom per picosecond within $0.1\;eV$ of the Fermi level. Thus, as long as $T_{10}$ is on the order of $\sim1\times10^{-4}$ there will be plenty of attempts to result in a single transmission event.

\begin{figure}[tb]
	\includegraphics[width=8.0 cm]{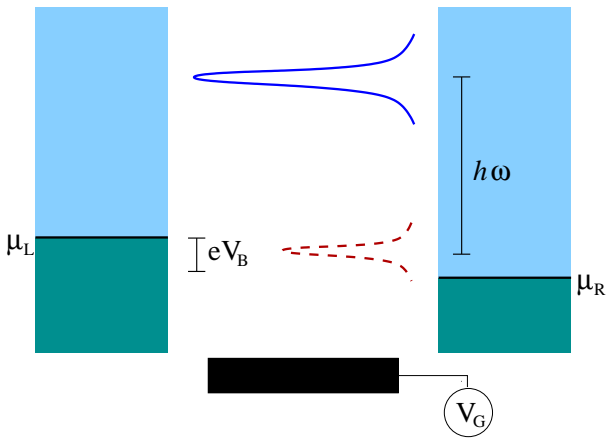} 
\caption{The principle of single electron transmission. A small bias is applied such that the bias window just covers the resonance: $eV_B\sim\Gamma$ and a gate voltage is tuned such that the resonance is located at: $\varepsilon_0\sim\mu_L-eV_B/2+\hbar\omega$. In the vibrational ground state the transmission function $T_{00}$ (solid line) is zero in the bias window. Exciting the first vibrational state changes the transmission function which is dominated by the inelastic part $T_{10}$ (dashed line) in the bias window. A vibrational excitation of the molecule will thus result in a single electron being transmitted.}
\label{fig:bias}
\end{figure}

The setup is illustrated in Fig. \ref{fig:bias} where a small bias voltage $V_B\sim\Gamma/e$ has been applied and a gate voltage $V_G$ has been tuned such that the position of the resonance is located $\hbar\omega$ above the bias window. It is crucial that the electronic resonance has a width much smaller than the quantum of vibration ($\Gamma\ll\hbar\omega$), since otherwise there will be a small but constant transmission probability when the molecule is in the vibrational ground state.

\begin{figure}[tb]
	\includegraphics[width=7.0 cm]{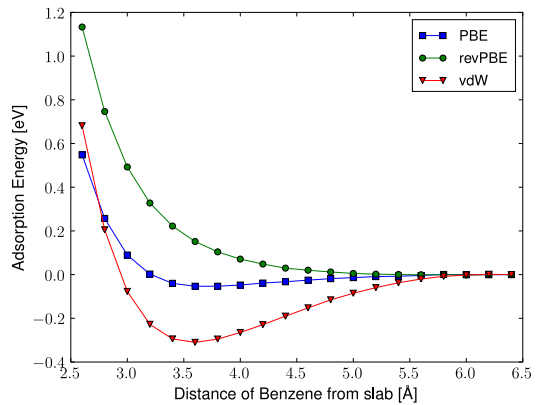}
\caption{Adsorption energy of benzene on Au(111) as a function of distance to the surface, calculated with three different functionals. The molecular states are weakly hybridized with the metallic states and only the functional including the van der Waals interaction gives the correct adsorption well.}
\label{fig:pes}
\end{figure}

\begin{figure}[tb]
	\includegraphics[width=5.0 cm]{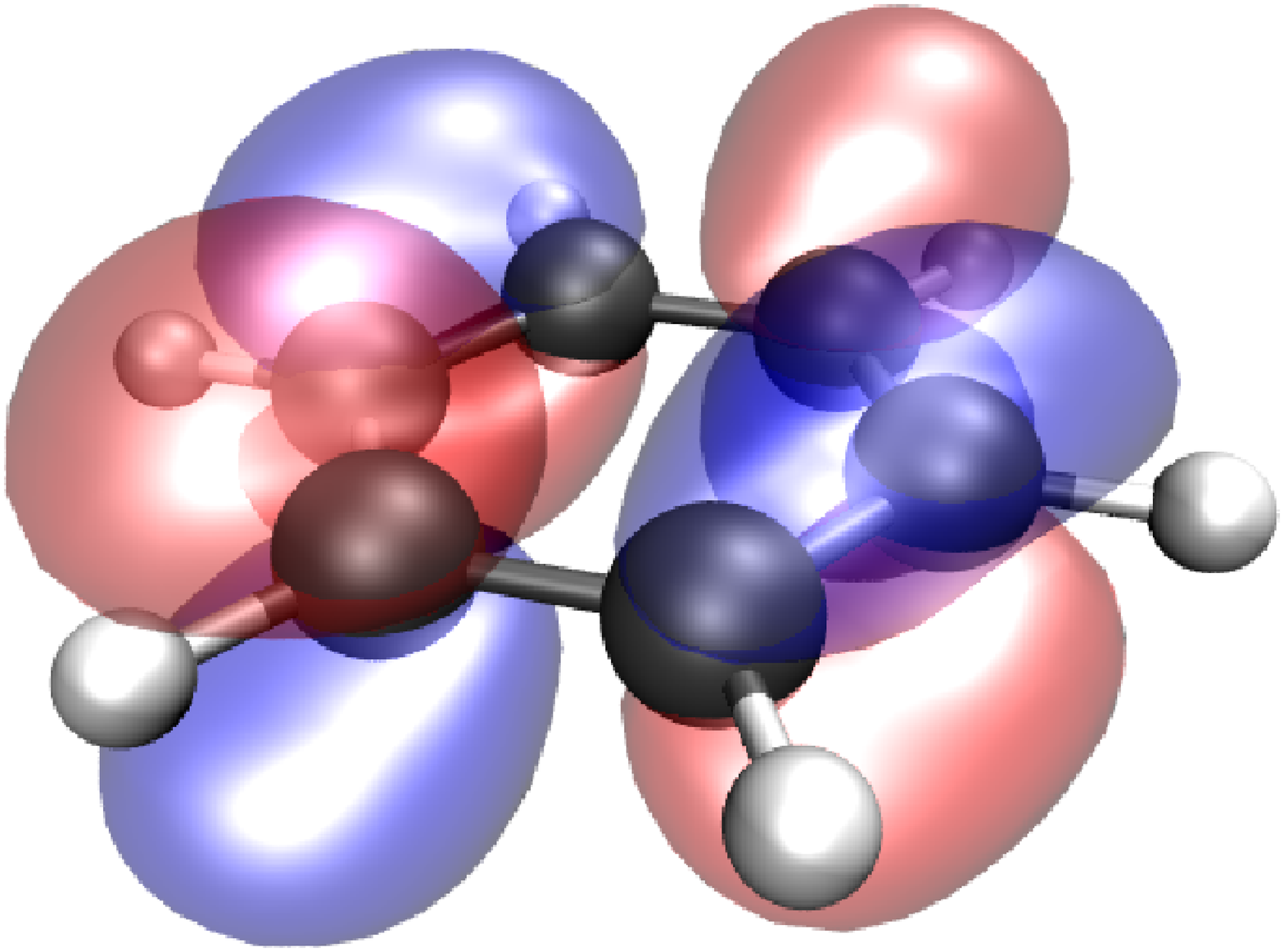}
	\includegraphics[width=3.0 cm]{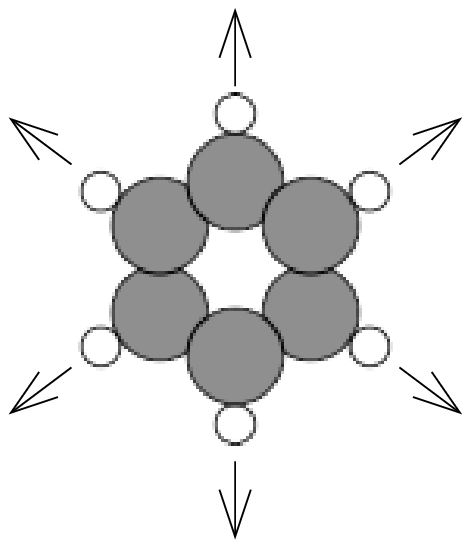}
\caption{Left: The lowest unoccupied orbital (LUMO) of benzene. The red and blue surfaces are positive and negative isosurfaces of the wavefunction. Right: The vibrational mode of highest energy which can be excited by a transient occupation of the LUMO.}
\label{fig:benzene}
\end{figure}

As an example we have performed a density functional theory (DFT) study of a benzene molecule sandwiched between two Au(111) contacts. The calculations were performed with the code \texttt{gpaw} \cite{gpaw,mortensen} which is a real-space DFT code using the projector augmented wave method \cite{blochl1}. The contact were simulated by a three layer Au(111) slab where the the top layer has been relaxed. We used a supercell with 12 Au atoms in each slab layer which were sampled by a $4\times4$ grid of K-points and $12.2\;A$ of vacuum. Benzene were added with its plane parallel to the surface and the adsorption energy was calculated as a function of distance to the slab. This is shown in Fig. \ref{fig:pes} for PBE \cite{PBE}, revPBE \cite{revPBE} and vdW \cite{dion} functionals. The PBE and revPBE functional shows weak and no bonding respectively whereas the vdW functional shows a $0.3\;eV$ minimum at $3.6\;A$.
The weak van der Waals bonding indicates that benzene is physisorbed on Au(111) which means that the molecular orbitals are weakly hybridized with metallic states as required by vibrationally mediated transmission of single electrons. This can also be seen explicitly from the Kohn-Sham projected density of states from which we estimate the width of HOMO and LUMO resonances to be $\Gamma_L\sim0.01\;eV$. It should be noted that we assumed a LUMO state in Eq. \eqref{H} and Fig. \ref{fig:transmission}, but the analysis is equally valid for transmission of hole states mediated by the HOMO and in the following we will consider both  types of resonance. Benzene has two degenerate HOMOs and two degenerate LUMOs which have the potential to mediate transmission of electrons through the molecule. One of the LUMOs is shown in Fig. \ref{fig:benzene} and it is expected that a transient occupation of the orbital may induce internal forces in the molecule and thus couple to the vibrational modes of the molecule. We have performed a DFT-based normal mode analysis of the benzene molecule which has 36 vibrational modes. There are 6 degenerate highly energetic modes with $\hbar\omega_i\sim0.39\;eV$ and the rest of the modes are evenly distributed in the interval $\hbar\omega_i\sim0-0.20\;eV$. The high energy modes involve the hydrogen atoms oscillating in the plane of the molecule along the individual H-C bonds as shown in Fig. \ref{fig:benzene}. The HOMO and LUMO states are expected to couple to several of the molecular vibrational states but we will focus on the high energy modes which have highly separated vibrational sidebands in the weak coupling limit. The coupling constants $\lambda_i$ can be related to the excited state potential energy surface $V_a$ associated with the LUMO being occupied or the HOMO being emptied\cite{olsen1}:
\begin{align}\label{lambda}
\lambda_i=\frac{l_i}{\sqrt{2}}\frac{\partial V_a}{\partial u_i}\bigg|_{u_i=u_i^0},\qquad l_i=\sqrt{\frac{\hbar}{m_i\omega_i}},
\end{align}
where $u_i$ is the coordinate of the $i$'th normal mode, $u_i^0$ its equilibrium position in the electronic ground state, $m_i$ its effective mass and $V_a$ the excited state potential energy surface associated with the resonance \cite{gavnholt, olsen1}. To obtain $V_a$ we have used the method of linear expansion $\Delta$ self-consistent field DFT \cite{gavnholt} which allows us to calculate the excited state energies while moving the atoms along the high energy mode where the 6 hydrogen atoms move in phase (see Fig. \ref{fig:benzene}). The results for this mode are $\lambda^{HOMO}=27\;meV$ and $\lambda^{LUMO}=9\;meV$ which should be compared with the quantum of oscillation $\hbar\omega_i\sim0.39\;eV$. The coupling is thus rather weak and we obtain maximum probabilities of inelastic transmission of $5\times10^{-3}$ and $5\times10^{-4}$ respectively (with $\Gamma=0.01\;eV$) at the lower vibrational sideband. However, assuming the vibrational lifetime to be on the order of nanoseconds, the probabilities are most likely large enough that an elastic transmission event will occur. Furthermore, a small coupling constant means that the ratio $T_{11}/T_{10}$ becomes very small and vibrational excitation will thus nearly always result in only a single electron being transmitted.

In summary, we have presented a method, which allows one to control the transmission of single electrons in weakly coupled molecule-metal junctions. The transmission is mediated by exciting a vibrational mode of the molecule while a gate voltage is tuned such that the resonant state is kept a quantum of vibrational energy above the bias window. It is assumed that such an excitation can be obtained with an external perturbation, for example a short laser pulse. The requirement of weak metallic coupling ($\Gamma\ll\hbar\omega$) is essential since it excludes elastic transmission in the vibrational ground state. For small vibrational coupling ($\lambda\ll\hbar\omega$) the junction will then be highly reliable and always give rise to one electron being transmitted. For large vibrational coupling $(\lambda>\hbar\omega)$ a vibrational excitation will typically result in a few electrons being transmitted due to a non-vanishing elastic transmission. To illustrate a quantitative approach to obtain the parameters of a real system, we have used DFT to calculate coupling parameters for a benzene molecule interacting with two gold contacts. Since benzene is bound by van der Waals forces and only show a weak hybridization with metallic states it satisfies the minimum requirement for the principle to work. However, there may be others reasons why this system is not well suited for experiments of this kind and it would be very interesting to investigate the principle in systems where transmission through single molecules with significant vibrational coupling has been observed \cite{park, yu, pasupathy, sapmaz, leturcq}.

We are grateful to K. S. Thygesen for advice and comments on the manuscript. This work was supported by the Danish Center for Scientific Computing. The Center for Individual Nanoparticle Functionality (CINF) is sponsored by the Danish National Research Foundation.

\appendix
\section{}
In this appendix we will show the details of the calculations leading to Fig. \ref{fig:transmission} for a single vibrational mode of frequency $\omega_0$ and coupling $\lambda_0$. Within scattering theory the transmission matrix $T_n(\varepsilon_i,\varepsilon_f)$ for a vibrationally excited state $n$ can be expressed as\cite{wingreen}
\begin{align}\label{trans}
 T_n(\varepsilon_i,\varepsilon_f)=&\Gamma_L\Gamma_R\int\frac{d\tau dtds}{2\pi\hbar^3}e^{i[(\varepsilon_i-\varepsilon_f)\tau+\varepsilon_ft-\varepsilon_is]/\hbar}\notag\\
&\times G_n(\tau,s,t),
\end{align}
where
\begin{align}
G_n(\tau,s,t)=\theta(s)\theta(t)\langle n|c_a(\tau-s)c_a^\dag(\tau)c_a(t)c_a^\dag|n\rangle
\end{align}
is the two particle Green function of the vibrational state $n$. The Green function can be evaluated exactly in the wide band limit and the result is\cite{olsen2}
\begin{align}
G_n(\tau,s,t)=&G_R^0(t)G_R^{0*}(t)e^{ig\omega_0(t-s)}\notag\\
&\times e^{-gf_{\tau,s,t}}L_n[g(f_{\tau,s,t}+f_{\tau,s,t}^*)],
\end{align}
where $L_n$ is the $n$'th Laguerre polynomial, $g=\lambda_0^2/(\hbar\omega_0)^2$, $$G_R^0(t)=-i\theta(t)e^{-i(\varepsilon_0-i\Gamma/2)t/\hbar},$$ and 
\begin{align}
f_{\tau,s,t}=2-e^{-i\omega_0t}-e^{i\omega_0s}+e^{-i\omega_0\tau}(1-e^{i\omega_0t})(1-e^{i\omega_0s}).\notag
\end{align}
An explicit result for $T_n$ can be obtained by perfoming the integrals in Eq. \eqref{trans} after a Taylor expansion of $e^{-gf_{\tau,s,t}}$.

The result for the vibrational ground state $T_0$ involves $L_0(x)=1$ and has been calculated previously.\cite{wingreen} Here we simply state the result which is
\begin{widetext}
\begin{align}\label{T_0}
 T_0(\varepsilon_i,\varepsilon_f)=\Gamma_L\Gamma_Re^{-2g}\sum_{m=0}^\infty\frac{g^m}{m!}\delta(\varepsilon_i-\varepsilon_f-m\hbar\omega_0)\bigg|\sum_{j=0}^m(-1)^j\binom{m}{j}\sum_{l=0}^\infty\frac{g^l}{l!}\Big(\frac{1}{\varepsilon_i-\varepsilon_0-(j+l-g)\hbar\omega_0+i\Gamma/2}\Big)\bigg|^2.
\end{align}
It is clear that integrating over final state energies simply produces a sum over vibrational transitions such that the $n$'th term in Eq. \eqref{T_0} represents $T_{0n}$. Calculating $T_1(\varepsilon_i,\varepsilon_f)$ is a bit more involved since the integrand in \eqref{trans} now includes the first Laguerre polynomial $L_1(x)=1-x$. We start by writing
\begin{align}
T_1(\varepsilon_i, \varepsilon_f)=T_0(\varepsilon_i, \varepsilon_f)+\widetilde{T}(\varepsilon_i, \varepsilon_f),
\end{align}
where
\begin{align}
\widetilde{T}(\varepsilon_i, \varepsilon_f)=\Gamma_L\Gamma_Re^{-2g}g&\int_0^\infty dse^{i(\varepsilon_0-\varepsilon_i-g\omega_0+i\Gamma/2)s/\hbar}\exp(ge^{i\omega_0 s})\notag\\
  \times&\int_0^\infty dt e^{-i(\varepsilon_0-\varepsilon_i-g\omega_0-i\Gamma/2)t/\hbar}\exp(ge^{-i\omega_0 t})e^{-i(\varepsilon_i-\varepsilon_f)t/\hbar}\notag\\
 \times&\int_{-\infty}^\infty \frac{d\tau}{2\pi\hbar^3} e^{i(\varepsilon_i-\varepsilon_f)\tau/\hbar}\exp\Big[ge^{-i\omega_0\tau}e^{i\omega_0 t}(1-e^{-i\omega_0t})(1-e^{i\omega_0s})\Big]\notag\\
 \times&\Big[(1-e^{i\omega_0t})(e^{-i\omega_0t}-1)+(1-e^{i\omega_0s})(e^{-i\omega_0s}-1)\notag\\
 &-e^{-i\omega_0\tau}(1-e^{i\omega_0t})(1-e^{i\omega_0s})-e^{i\omega_0\tau}(1-e^{-i\omega_0t})(1-e^{-i\omega_0s})\Big].\notag
\end{align}
Taylor expanding the second exponential in the $\tau$ integral and performing the integration gives
\begin{align}
\widetilde{T}(\varepsilon_i, \varepsilon_f)=\Gamma_L\Gamma_Re^{-2g}\frac{g}{\hbar^2}&\int_0^\infty dse^{i(\varepsilon_0-\varepsilon_i-g\omega_0+i\Gamma/2)s/\hbar}\exp(ge^{i\omega_0s})\notag\\
  \times&\int_0^\infty dt e^{-i(\varepsilon_0-\varepsilon_i-g\omega_0-i\Gamma/2)t/\hbar}\exp(ge^{-i\omega_0t})\notag\\
 \times\bigg[&\sum_{m=0}^\infty\frac{g^m}{m!}e^{i\omega_0t}(1-e^{-i\omega_0t})^{m+2}(1-e^{i\omega_0s})^m\delta(\varepsilon_i-\varepsilon_f-m\hbar\omega_0)\notag\\
 &+\sum_{m=0}^\infty\frac{g^m}{m!}e^{-i\omega_0s}(1-e^{-i\omega_0t})^m(1-e^{i\omega_0s})^{m+2}\delta(\varepsilon_i-\varepsilon_f-m\hbar\omega_0)\notag\\
 &+\sum_{m=0}^\infty\frac{g^m}{m!}(1-e^{-i\omega_0t})^{m+1}(1-e^{i\omega_0s})^{m+1}\delta(\varepsilon_i-\varepsilon_f-(m+1)\hbar\omega_0)\notag\\
 &+\sum_{m=0}^\infty\frac{g^m}{m!}e^{-i\omega_0(s-t)}(1-e^{-i\omega_0t})^{m+1}(1-e^{i\omega_0s})^{m+1}\delta(\varepsilon_i-\varepsilon_f-(m-1)\hbar\omega_0)\notag\bigg].
\end{align}
The first two terms are each others complex conjugated and the last two terms factorizes (s and t integrals) into complex conjugated and the integrals can then be performed. The final result is rather complicated but consist of an infinite number of terms, each of which involves a delta function $\delta(\varepsilon_i-\varepsilon_f-m\hbar\omega_0)$, where $m$ runs from -1 to infinity. We can thus obtain $T_{10}$ and $T_{11}$ by collecting terms involving $m=-1$ and $m=0$ respectively. The results are
\begin{align}
T_{10}(\varepsilon_i)=\Gamma_L\Gamma_Re^{-2g}g\bigg|\sum_{l=0}^\infty\frac{g^l}{l!}\frac{1}{\varepsilon_i-\varepsilon_0-(l-1-g)\hbar\omega_0+i\Gamma/2}-\sum_{l=0}^\infty\frac{g^l}{l!}\frac{1}{\varepsilon_i-\varepsilon_0-(l-g)\hbar\omega_0+i\Gamma/2}\bigg|^2,\notag
\end{align}
and
\begin{align}
T_{11}(\varepsilon_i)=&\Gamma_L\Gamma_Re^{-2g}\bigg|\sum_{l=0}^\infty\frac{g^l}{l!}\frac{1}{\varepsilon_i-\varepsilon_0-(l-g)\hbar\omega_0+i\Gamma/2}\bigg|^2\notag\\ 
+2g&\Gamma_L\Gamma_Re^{-2g}\mathtt{Re}\bigg(\sum_{j=0}^{2}(-1)^j\binom{2}{j}\sum_{l=0}^\infty\frac{g^l}{l!}\frac{1}{\varepsilon_i-\varepsilon_0-(j+l-1-g)\hbar\omega_0+i\Gamma/2}\sum_{l=0}^\infty\frac{g^l}{l!}\frac{1}{\varepsilon_i-\varepsilon_0-(l-g)\hbar\omega_0-i\Gamma/2}\bigg)\notag\\
+g^2&\Gamma_L\Gamma_Re^{-2g}\bigg|\sum_{j=0}^{2}(-1)^j\binom{2}{j}\sum_{l=0}^\infty\frac{g^l}{l!}\frac{1}{\varepsilon_i-\varepsilon_0-(j+l-1-g)\hbar\omega_0+i\Gamma/2}\bigg|^2.\notag
\end{align}

\end{widetext}


\end{document}